\newcommand{\Cro}{{\rm Cro}}
\newcommand{\cro}{{\rm cro}}
\newcommand{\cI}{{\sl cI}}

\newcommand{\CI}{{\rm CI}}

\newcommand{\RecA}{{\rm RecA}}
\newcommand{\Ecoli}{{\sl E.~coli}}

\newcommand{\OR      }{$O_{\rm R}$}

\newcommand{\ORone   }{$O_{\rm R}1$}

\newcommand{\ORtwo   }{$O_{\rm R}2$}

\newcommand{\ORthree }{$O_{\rm R}3$}

\newcommand{\PR      }{$P_{\rm R}$}

\newcommand{\PRM     }{$P_{\rm RM}$}

\documentstyle[prl,aps,multicol,epsf,psfig]{revtex}

\begin{document}
\begin{title}
{Epigenetics as a first exit problem}
\end{title}
\vspace{0.5cm}

\author{E. Aurell$^{1,2,3}$ and K. Sneppen$^{3}$}
\vspace{0.5cm}
\address{
$^{1}$SICS, Box 1263, SE-164 29 Kista, Sweden \\
$^{2}$SANS/NADA/KTH, SE-100 44 Stockholm, Sweden \\
$^{3}$NORDITA, Blegdamsvej 17, DK-2100 Copenhagen, Denmark
}

\date{\today}

\maketitle

\vspace{0.5cm}

\centerline{Pacs numbers: 87,87.16-b}

\vspace{0.5cm}

\begin{abstract}
We develop a framework to discuss stability of epigenetic states
as first exit problems in dynamical systems with noise.
We consider in particular the stability of the lysogenic state of the
$\lambda$ prophage, which is known to exhibit exceptionally large stability.
The formalism defines a quantative measure of robustness of inherited
states.
\end{abstract}

\begin{multicols}{2}      

\vspace{0.5cm}

\begin{figure}[p]
%\vspace{-4mm}
%\mbox{
%	\epsfxsize=0.4\hsize
%	\hspace*{.18\hsize}
%	\epsffile{lfig010118.ps}
%}
%\centerline{\psfig{file=lfig010118.ps,angle=270,width=12cm,clip=}}
\centerline{\psfig{file=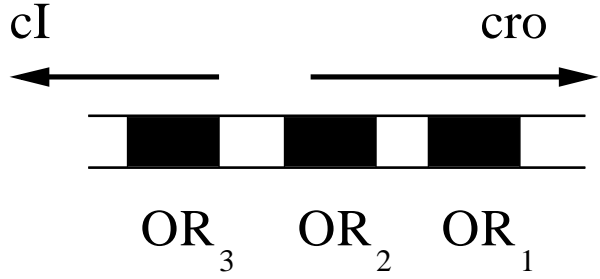,angle=270,width=8cm,clip=}}
%\epsffile{lfig010118.ps} 
\vspace{0.5cm}
\caption[]
{
Right operator complex, \OR, consisting of the three operators
\ORone, \ORtwo\  and \ORthree. \cI\ is transcribed when \ORthree\
is free and \ORtwo\ is occupied by \CI.
\cro\ is transcribed when both \ORtwo\ and \ORone\ is free.
\CI\ dimers bind cooperatively to \ORone\ and \ORtwo.
}
\label{f:OR}
\end{figure}

\begin{figure}[p]
%\vspace{-4mm}
\epsfxsize=0.6\hsize
\begin{center}
 \mbox{
	\epsffile{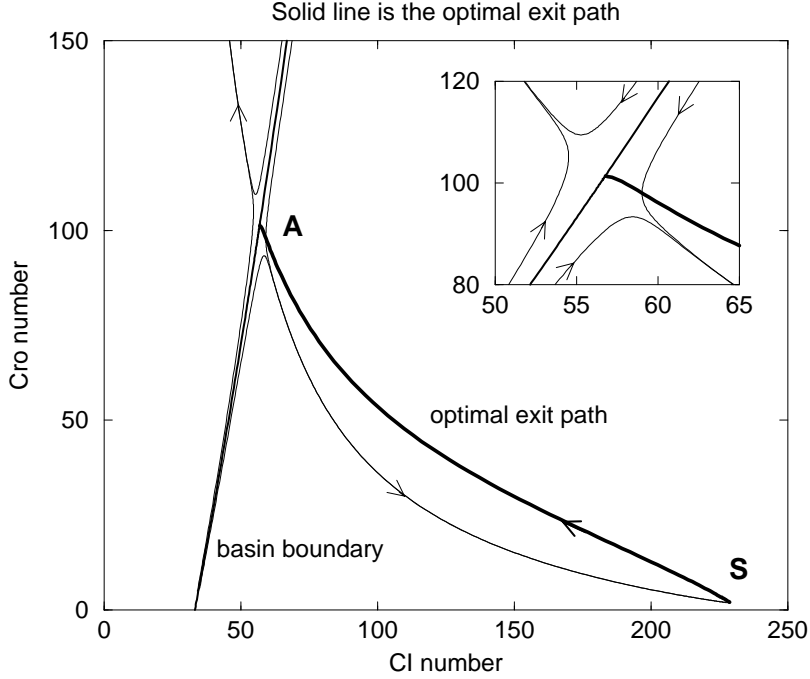}
}
\end{center}
%\centerline{\psfig{file=PhaseSpace.eps,width=10cm}}
%\epsffile{PhaseSpace.eps} 
%\vspace{0.5cm}
\caption[p]
{
The phase space plot of the dynamical system 
(\protect\ref{eq:coupled-equations}), and the optimal
exit path in the stochastic dynamical system (\protect\ref{eq:coupled-SDE}).
The lysogenic state is identified with a stable equilibrium $S$ at
$( N_{\CI},N_{\Cro}) \approx (225,1)$. The basin of attraction of this equilibrium
is delimited by a separatrix (basin boundary), which passes through the unstable
equilibrium point $A$ at $( N_{\CI},N_{\Cro}) \approx (56,101)$.
The separatrix crosses the \CI--axis at  $N_{\CI}\approx 33$.
Also indicated is the most probable exit path (full line with arrow)
from $S$ to $A$. 
Insert shows a blow-up around unstable equilibrium $A$.
Note that the most probable exit path goes into $A$ at a different angle
compared to the trajectories of 
(\protect\ref{eq:coupled-equations}) going out of $A$.
Parameter values are as in \protect\cite{ABJS}.
}
\label{f:phase-space}
\end{figure}

Epigenetics is concerned with inherited states in living systems,
which are not encoded as genes, but as the (inherited) patterns
of expressions of genes.
Modulation of gene expression, or functional genomics,
underlies a wide number of biological
phenomena, e.g. efficient use of nutrients available
to an organism at a particular time.
A familiar example of
inherited patterns of gene expression is
cell differentiation in multicellular organisms,
which, if once established, can be propagated for
long times.
The stability of epigenetic states is important,
as it simultaneously enables an organism to
maintain a favorable state, and to keep the
ability to change that state in a coordinated
manner, if external conditions so dictate.
\\\\
Some of the simplest examples of two-state systems in
biology are
found among bacteriophages, DNA viruses growing on bacterial hosts.
When in the state of
being stably integrated into the
genome of the host, known as lysogeny,
one set of viral genes is expressed.
When on the other hand the virus is performing other tasks,
such as directing translation of
viral proteins leading to lysis,
killing the bacterial host, other sets of genes are expressed.
The classical example is the lysogenic state
of phage $\lambda$ in {\it Escherichia coli} 
\cite{Johnson,Shea,Ptashne}.
Upon infection of an \Ecoli\ cell,
either $\lambda$ enters a path leading to lysis,
or it enters lysogeny, which can then
be passively replicated for very long
times.  Indeed, the wild-type rate of spontaneous loss of $\lambda$ lysogeny
is only about $10^{-5}$ per cell and generation~\cite{Ptashne},
a life-time of the order of five years.
Moreover, this number is but a consequence of random activation
of another part of the genetic system, the SOS response
involving RecA, and the intrinsic loss rate has in several
independent experiments been found to be less than
 $10^{-7}$ per cell and generation
\cite{Rozanov,Little,ABJS}, possibly as low as
$2 \cdot 10^{-9}$~\cite{Little-priv}.
The rate of mutations in the part of the lambda genome involved
in lysogeny is between $10^{-6}$ and  $10^{-7}$ per generation
\cite{Little,ABJS}.
Epigenetics is therefore in this system 
more stable than the genome itself. 
\\\\
One may recall that E. Schr\"odinger in
{\it ``What is life?''}\cite{Schrodinger}
starts by imagining that the stability of genetic inheritance
stems from dynamic equilibria involving macroscopic numbers of molecules. 
On the basis of the then
recent experimental data on the dependence of mutation rate on
radiation,
this hypothesis is discarded
in favor of a molecular model of genetic memory, fore-shadowing the
DNA-RNA machinery.
Epigenetics, in particular $\lambda$ lysogeny as
presented below, gives an example where Schr\"odinger's
first idea is essentially correct. The number of molecules needed to
achieve stability, over biologically relevant time-scales,
is however surprisingly small, only in the order
of hundreds, and thus rather in the mesoscopic than in the macroscopic
range. Such a number however nevertheless
gives fluctuations in gene expression
\cite{Arkin98}, and accordingly but a finite stability of many states. 
\\\\
A stable state can be likened to a control switch that is on.
For $\lambda$ the analogy is quite direct \cite{Johnson,Ptashne}: 
lysogeny is maintained by protein molecules and $\lambda$ DNA
around an operator \OR, which consists of three binding sites
\ORone, \ORtwo\ and \ORthree, overlapping with two promotor sites
\PRM\ and \PR, see Fig.~\ref{f:OR}.
At the binding sites either one of two regulatory proteins \CI\
and \Cro\ can bind. These proteins are produced from the
corresponding genes \cI\ and \cro, which abut \OR, and which
are regulated by \CI\ and \Cro.
Hence transcription of \cro\ starts at
\PR, which partly overlaps \ORone\ and \ORtwo, while
transcription of \cI\ starts at \PRM, which partly overlaps
\ORtwo\ and \ORthree.
The affinity to the two promotors of RNA polymerase, the enzyme which catalyses
the production of mRNA transcripts from DNA,
depends on
the pattern of \Cro\ and \CI\ bound to the operator sites.
The rates of production of the two proteins are therefore
functions of the concentrations of the proteins
themselves, and balance decay and dilution in a stable stationary state
with 200-300 \CI\ and, on the average, few \Cro\ per bacterium
\cite{Ptashne}.
This is a logical switch, because if
\CI\ concentration
becomes sufficiently low, the increased activation of \cro\
increases \Cro\ concentration and decreases \cI\ activation,
so that lysogeny is ended and lysis follows.
\end{multicols}
%----------------------------------------------------------
%-------------------------------------------
\begin{multicols}{2}
The simplest mathematical model which embodies Fig.~\ref{f:OR} is
a set of coupled equations for the time rate of change of numbers
of \CI\ and \Cro\ in a cell~\cite{ReinitzVaisnys}:
\begin{equation}
\label{eq:coupled-equations}
\begin{array}{lcl}
\dot N_{\CI} &=&  \phi_{\CI}(N_{\CI} , N_{\Cro} ) \\
\dot N_{\Cro} &=& \phi_{\Cro} (N_{\CI} , N_{\Cro} ) 
\end{array}
\end{equation}
where the net production rates are
\begin{equation}
\label{eq:phi-defs}
\begin{array}{lclcl}
\phi_{\CI} &=& f_{\CI }(N_{\CI },N_{\Cro }) &-& N_{\CI}/\tau_{\CI} \\
\phi_{\Cro} &=& 
f_{\Cro}(N_{\CI},N_{\Cro})& -& N_{\Cro}/\tau_{\Cro}
\end{array}
\end{equation}
The production terms $f_{\CI}$ and $f_{\Cro}$ are
functions of \CI\ and \Cro\ concentrations. With no \Cro\
in the system, the curve of $f_{\CI}$ {\it vs.} \CI\ concentration has
been experimentally measured~\cite{HawleyMcClure82}. 
As reviewed in \cite{ABJS}, these measurements
are consistent with 
the best available data on protein-DNA affinities~\cite{Kim87,Koblan92,Takeda92}, 
dimerization constants~\cite{Koblan91}, 
initiation rates of transcriptions of the
genes, and the efficiency of translation of the mRNA transcripts into
protein molecules. 
The decay constant $\tau_{\CI}$ is proportional to
the bacterial life-time, since \CI\ molecules are not actively degraded
in lysogeny, while $\tau_{\Cro}$ is about~30~\% smaller~\cite{Pakula}.
We remark that there is considerably more experimental
uncertainty in the binding of \Cro, both to other \Cro\ and
to DNA, than the binding of \CI, see e.g.~\cite{DarlingHoltAckers}.
As a minimal model of the switch, we take $\tau_{\CI}$ and 
$\tau_{\Cro}$ from data, and deduce $f_{\CI}$ and $f_{\Cro}$ at 
non-zero concentrations of both \CI\ and \Cro\ with a standard
set of assumed values of all binding constants, as done in \cite{ABJS}. 
Such a model is conveniently visualized by
the phase space plot in  Fig.~\ref{f:phase-space}.
\\\\
If system (\ref{eq:coupled-equations}) is in lysogeny, 
i.e. in the stable equilibrium $S$, it
will stay there indefinitely. The system leaves lysogeny when
external perturbations push \CI\ concentration to the left
of the separatrix. {\it In vivo}, as sketched above,
the important perturbation is \RecA--mediated
self-cleavage of \CI, as a by-product of the SOS DNA repair mechanism
when host DNA is damaged. The functional purpose of the switch,
for the virus, is hence to sense if the host is in danger, and,
if so, jump ship.
\\\\
If the numbers of \CI\ and \Cro\ were macroscopically large,
then (\ref{eq:coupled-equations}) would be an entirely accurate
description of the dynamics. The numbers
are however
only in the range of hundreds. The actual production process
is influenced by many chance events, such as the time it takes
for a \CI\ or a \Cro\ in solution to find a free operator site,
or the time it takes a RNA polymerase molecule to find and attach
itself to an available promotor. If in a time interval $\Delta t$
the expected number of molecules produced is $f\Delta t$, 
then the number produced in an actual realization 
has scatter $\sqrt{f\Delta t}$.
As a minimal model of the switch with finite-$N$
noise, we therefore consider the following system of
two coupled stochastic differential equations, with two independent
standard Wiener noise sources $(d\omega^{\CI}_t, d\omega^{\Cro}_t)$:
\begin{equation}
\label{eq:coupled-SDE}
\begin{array}{lclcl}
d N_{\CI} &=& \phi_{\CI}\, dt 
&+& g_{\CI} \, d\omega^{\CI}_t \\
d N_{\Cro} &=& \phi_{\Cro}\, dt §
&+& g_{\Cro} \, d\omega^{\Cro}_t
\end{array}
\end{equation}
We assume that there is
an equal amount of finite-$N$ noise in decay as in production, and the
two noise amplitudes are hence
\begin{equation}
\label{eq:noise-amplitudes}
\begin{array}{lcl}
g_{\CI}  &=& \sqrt{f_{\CI} + N_{\CI}/\tau_{\CI}}\\
g_{\Cro} &=& \sqrt{f_{\Cro}+N_{\Cro}/\tau_{\Cro}}
\end{array}
\end{equation} 
\\\\
The problem of escape from a stable equilibrium point like $S$ under
a dynamics like (\ref{eq:coupled-SDE}) is a first-exit problem in
the theory of stochastic processes. As such it is solved in Wentzel-Freidlin
theory~\cite{WentzellFreidlin,MaierStein}. A special case of Wentzel-Freidlin
theory is well-known from chemical physics, namely if the noise
amplitudes $(g_{\CI},g_{\Cro})$ are constant and the drift field
is a potential field. If so, the problem of escape from $S$ is
Kramers' classical problem of thermal escape from a potential well~\cite{Kramers,Hanggi}.
The more general Wentzel-Freidlin problem has
both similarities and differences to Kramers' problem, as we will now explain. 

\end{multicols}

%----------------------------------
%----------------------
\begin{multicols}{2}
Proceeding heuristically, we note that the probability of a given realization of
the noise in time [0,T] is
\begin{equation}
\label{eq:probability}
\begin{array}{l}
\hbox{Prob}(\{\omega^{\CI}_t, \omega^{\Cro}_t\}_0^T) \\
\quad \propto \quad
\exp\left(-\frac{1}{2}\int_0^T (\dot\omega^{\CI}_t)^2 + (\dot\omega^{\Cro}_t)^2\, dt \right)
 \\
\quad = \quad \exp\left(-\int_0^T \frac{(\dot N_{\CI} - \phi_{\CI} )^2}{\Gamma_{\CI}} +
\frac{(\dot N_{\Cro} - \phi_{\Cro} )^2}{\Gamma_{\Cro}}
\, dt \right)
\end{array}
\end{equation}
where we have introduced the diagonal elements of the
diffusion matrix, 
$\Gamma_{\CI}=g_{\CI}^2$ and $\Gamma_{\Cro}=g_{\Cro}^2$. 
\\\\
Of all the realizations that move the system from $S$ to $A$, the
most probable is the one that minimizes the action functional
\begin{equation}
{\cal A} = \frac{1}{2}
\int_0^T \left(\frac{(\dot N_{\CI} - \phi_{\CI} )^2}{\Gamma_{\CI}} +
\frac{(\dot N_{\Cro} - \phi_{\Cro} )^2}{\Gamma_{\Cro}}
\right)\, dt
\label{eq:action}
\end{equation}
where the initial position is $S$, the final position $A$, and the
minimization is taken over all paths that go from $S$ to $A$
in time $T$.
If ${\cal A}^{min} >> 1$ it can be proved (see~\cite{MaierStein})
that the most probable exit point
from the basin of attraction of $S$ is indeed
$A$, and the rate of exit is to leading order 
\begin{equation}
\hbox{Rate}(\hbox{exit})
\propto \exp\left(-{\cal A}^{\hbox{min}}\right)
\label{eq:probability-exit}
\end{equation}
The asymptotic 
formula (\ref{eq:probability-exit})
also contains a prefactor of dimension one over 
time~\cite{WentzellFreidlin,MaierStein}, which
in the case of a potential field reduces to the prefactor in 
Kramers' formula~\cite{Kramers}. 
In our case the prefactor is of order once per bacterial generation.
The optimal exit path obeys the appropriate
Euler-Lagrange equations, being the extremal 
of variations of ${\cal A}$.
Since the Lagrangian in (\ref{eq:action}) is not explicitly time-dependent, 
the Hamiltonian
\begin{equation}
{\cal H} = \frac{1}{2}\left(\Gamma_{\CI}\, p_{\CI}^2
+  \Gamma_{\Cro} \,p_{\Cro}^2\right) +  p_{\CI}\phi_{\CI} + p_{\Cro}\phi_{\Cro}
\label{eq:hamiltonian}
\end{equation}
is conserved along the path. The momenta $(p_{\CI},p_{\Cro})$ are
conjugate to the generalized coordinates $(N_{\CI},N_{\Cro})$, and
the Euler-Lagrange equations are
equivalent to Hamilton's equations
$\dot{\vec{N}} = \frac{\partial{\cal H}}{\partial \vec p}$ and
$\dot{\vec{p}} = -\frac{\partial{\cal H}}{\partial \vec N}$.
We note that in this auxiliary
classical mechanical system, the diffusion constants play
the role of space-dependent elements of an inverse mass matrix,
while the drift field is somewhat similar to a magnetic potential.
We also note that the energy of the optimal exit path, the
value of ${\cal H}$ along that path, must be
non-negative, since the drift field vanishes at the two end points.
On the other hand, we have in general $\partial {\cal A}/\partial T = -E$,
where $E$ is the energy and $T$ is the transit time. It hence follows
that the optimal exit path is a zero-energy path from $S$ to $A$
under the Hamiltonian in (\ref{eq:hamiltonian}).
\\\\
Fig~\ref{f:phase-space} shows the optimal fluctuation path.
In contrast to thermal escape from a potential
well, where the optimal path is always opposite in direction
and equal in size to the drift field, there is no obvious
simple prescription to directly compute the path
from $S$ to $A$ in
Fig.~\ref{f:phase-space}. 
The Hamiltonian analogy
however suggest the following numerical
procedure, using the relaxation method of
computing solutions to 2-point boundary problems
in an ODE~\cite{NumericalRecipies,OtherRef}.
\\\\
We first find a natural parameter in the system,
typically one of the binding constants,
and vary that to get close to the bifurcation where
the stable and unstable equilibria ($S$ and $A$)
coalesce. The diffusion parameters $\Gamma_{\CI}$ and
$\Gamma_{\Cro}$ are then practically constant
in a neighbourhood around both points, while the drift field
is small. We can then
compute a path between the two points at high
energy (equivalently, at a small transit time $T$), 
starting from a trial solution, which is a straight path at constant
speed. In other words, $\dot{\vec N}$ is taken constant along the trial path,
and ${\vec p} = \frac{\partial {\cal L}}{\partial\dot{\vec N}}$
is given by $(\Gamma^{-1})\cdot\dot{\vec N} - \vec\phi$. 
The energy is then lowered
incrementally, and the optimal path at each energy found by relaxation,
using the previous solution as the trial solution at the new
energy. A zero-energy path can thus be found close to the bifurcation,
and, by changing back the parameter in small steps, 
again using relaxation to find each new path,
a zero-energy path can be found at the original parameter value. 
The zero-energy motion in the intermediate neighbourhoods
of the two points always needs to be taken care of by a local calculation,
as explained in~\cite{MaierStein}.
%----------------------------------------------------------
\end{multicols}
\begin{figure}[p]
\vspace{-4mm}
\epsfxsize=0.4\hsize
\hbox{
	\epsfxsize=0.4\hsize
	\epsffile{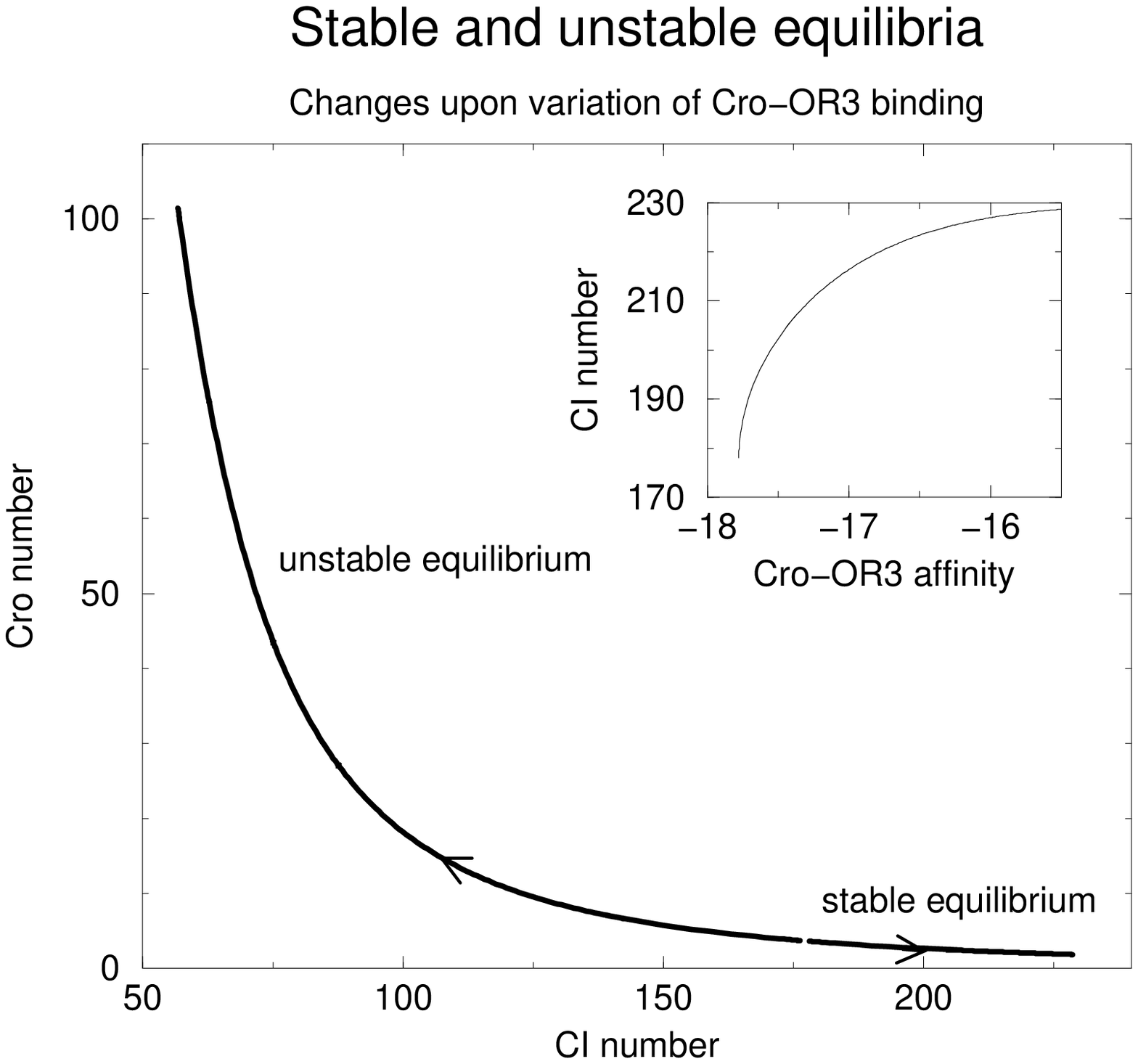}
	\hspace{1.0cm}
	\epsfxsize=0.4\hsize
	\epsffile{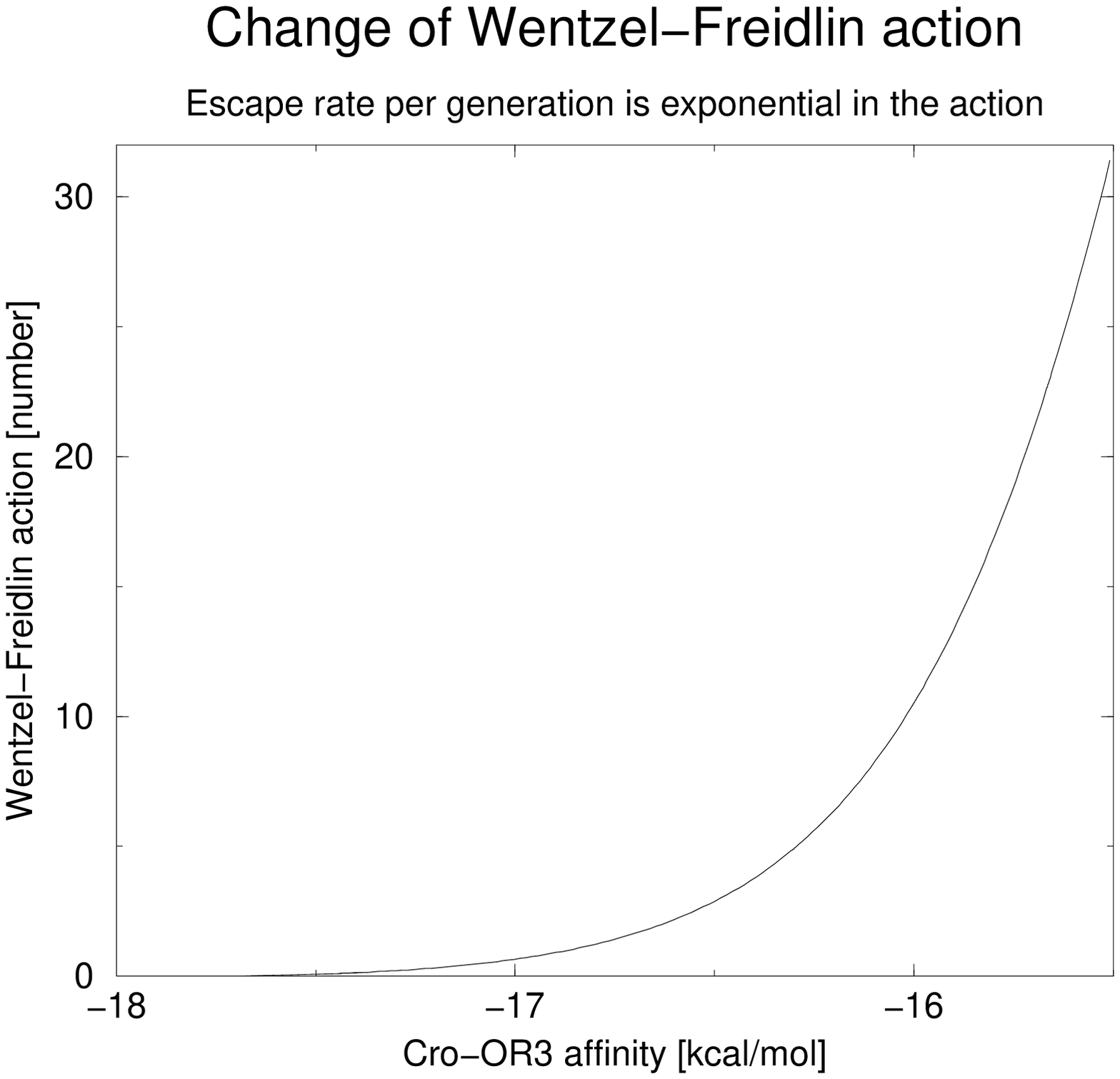}
}
%\centerline{\psfig{file=Points.eps,angle=270,width=12cm,clip=}} 
%\centerline{\psfig{file=Points.eps,width=6cm}\hspace{1.0cm}\psfig{file=Action.eps}} 
%\centerline{\epsffile{Points.eps}\hspace{1.0cm}\epsffile{Action.eps}} 
%\hspace{1.0cm}
%\epsfxsize=0.4\hsize
%\hbox{
%	\hspace*{.18\hsize}
%	\epsffile{Action.eps}
%}
%\centerline{\psfig{file=Action.eps,angle=270,width=12cm,clip=}} 
%\centerline{\psfig{file=Action.eps,width=6cm}}
%\centerline{b} 
\centerline{{\bf a}\hspace{7.0cm}{\bf b}} 
\vspace{0.5cm}
\caption[]
{
Systemic changes due to changes in affinity of \Cro\
to operator site \ORthree.
The standard value is $-15.5\,\hbox{kcal/mol}$. Stronger 
binding energies are investigated for use in numerical procedure
(see main text), and to explore robustness of lysogeny to
parameter changes.
Other parameters are as in \protect\cite{ABJS}.
{\bf a)} Location of stable equilibrium ($S$) and
unstable equilibrium ($A$) as affinity is varied. 
The two equilibria are born in a bifuraction
at affinity $-17.78\,\hbox{kcal/mol}$
and $(N_{\CI},N_{\Cro})\approx (176,3.73)$. At increasing
value of
binding energy (weaker binding), the equilibria move
apart, as indicated by the arrows. Insert shows the $N_{\CI}$
value of the lysogenic state (stable equilibrium), as function of
the affinity, in $\hbox{kcal/mol}$. 
{\bf b)} Wentzel-Freidlin action as function of affinity.
The escape rate from lysogeny is
exponential in the action, with a prefactor of the order
once per bacterial generation. At the standard value (-15.5 \hbox{kcal/mol}),
the predicted rate from the model
is hence about once in $10^{13}$ generations.
We note that the escape rate depends very sensitively on parameters.
A change of affinity by  $1\,\hbox{kcal/mol}$ to 
$-16.5\,\hbox{kcal/mol}$ gives an action of about $2.5$,
and an expected lifetime of the lysogenic state of only about
ten generations.
}
\label{f:dependence}
\end{figure}
%--------------------------------------------
\begin{multicols}{2}
There is an emerging consensus in
molecular biology and 
biological physics that 
chemical networks in living cells have to
be robust~\cite{Savageau,Leibler,Bialek}.
That is, they have not only
to work under some conditions, but should work under a wide 
variety of conditions, possibly even under change or replacement 
of parts of the network, see e.g. recent
mathematical modelling of cell signaling~\cite{BhallaIyengar}, and  
of a genetic control network for cell differentiation in
{\it Drosophila}~\cite{vonDassow}.
For the $\lambda$ phage, robustness of lysogeny has been 
experimentally established
for several large modifications of the \OR\ complex \cite{Little}.
The present work allows us to quantify robustness
of epigenetic states.
A state only exists at all
if deterministic equations like
(\ref{eq:coupled-equations}) have a stable
equilibrium with the corresponding properties.
This state is stable for long times, 
even if the number of molecules
involved is small, if the action ${\cal A}$
in (\ref{eq:action}) and (\ref{eq:probability-exit})
is much larger than unity.
The state is finally robust, if under a
typical change of 
a changeable parameter $\mu$, the state still
exists and is stable.
This means that
$\Delta \mu \cdot \frac{d{\cal A}}{d\mu}$
must be significantly less than ${\cal A}$,
where the typical change
$\Delta \mu$ could be the
change in binding energy upon
a single point mutation, of order
$1\,\hbox{kcal/mol}$.
\\\\
In Figs.~\ref{f:dependence}\,a,b) we examine
lysogenic stability as function of one parameter, 
the binding of \Cro\ to \ORthree. 
If we first disregard the noise, we see that a change 
of affinity by $2.25\,\hbox{kcal/mol}$ brings the
stable and§ unstable equilibria together, such that the
lysogenic state disappears altogether. 
We also observe a sensitive dependence of the position of the
unstable equilibrium, while the number of \CI\ in the
stable equilibrium (lysogenic state) only changes by 30 \%.
The lysogenic state therefore looks qualitatively similar
over this range of parameters. 
These are features
of the model embodied by equations~(\ref{eq:coupled-equations}) only.
If we then bring in our model of the noise,
equation~(\ref{eq:coupled-SDE}),
we see that the action ${\cal A}$ changes from
more than 30 to less than 3 when affinity changes
by $1\,\hbox{kcal/mol}$, the approximate change
of binding energy under a single point mutation.
Such a change hence suffices to destabilize the switch
over biologically relevant time-scales. and the
model is therefore not robust
to such changes, in contradiction to~\cite{Little}. This implies 
the presence of some additional mechanism, in order
for robustness to prevail. 
We stress that this lack of robustness is an inherent property
of the model, true for all variations of parameters
that have been put forward to quantitatively describe
these generally accepted mechanisms of 
control, including the recent report that 
\Cro\ may in fact bind cooperatively to \ORone~\cite{DarlingHoltAckers}.
\\\\
In conclusion,
we have examined the general problem
of excape from a stable equilibrium in more than one dimension,
and demonstrated how this determines the 
stability of states of genetic networks.
In contrast to Kramers' escape from a potential well,
the stability of inherited states in such networks
is not a mathematically and computationally
trivial problem. The most
likely exit path does not go along a steepest decent
of a potential -- there is no potential. Instead, 
such a path can be
described as a zero-energy trajectory 
between two equilibria in an auxiliary classical mechanical system.
Finding it involves similar numerical problems as e.g. computing heteroclinic
orbits in celestial mechanics.
The overall lesson of this study
is that an examination of equilibria 
and their bifurcations with changing parameter values
allow us to quantify both the stability
and the robustness of particular states of 
a genetic control system.
\\\\
\section*{Acknowledgements}
\label{s:acknowledgements}
We thank Prof. G. Dahlquist for suggesting the
numerical procedure used to find the minimum of the
action, and B.~Altshuler, S.~Brown, P.~Kraulis,
P.~Muratore-Ginanneschi and B.~\"Obrink
for discussions and valuable comments.
E.A. thanks
the Swedish Natural Research Council for support under grant
M-AA/FU/MA 01778-333.
K.S. thanks the ITP (U.C. Santa Barbara)
for hospitality, and consequently the
National Science Foundation for
financial support under grant No. PHY99-07949.

%\begin{thebibliography}{10}

\end{multicols}

\end{document}